\documentclass[twocolumn,preprintnumbers,amsmath,amssymb]{revtex4}
\usepackage{graphicx}
\usepackage{dcolumn}
\usepackage{bm}
\begin{document}
\title{\bf Comments on `Irreversibility in Response to Forces Acting on Graphene Sheets'}
\author{Mehdi Neek-Amal ~\footnote{neekamal@srttu.edu} }
\affiliation{Department of Physics, Shahid Rajaee University,
Lavizan, Tehran 16785-136, Iran.}

\maketitle

In Ref.~[1] the
compression-relaxation  mechanism (CRM) for a graphene sheet (GS) was reported
to be irreversible and resulted in static ripples on GS, such that for T $<$Tc the free-energy of the rippled
GS is smaller than that of roughened GS. We will point
out several technical difficulties, such as the use of the relation
$\Delta A=\langle W\rangle$ for the free energy calculations  and the
definition of the rough state, with their simulations~\cite{Fig.1}. We show
that (at T $<$Tc) their introduced rough state suffer boundary stress, thus
is a rippled state and their obtained inequality $A_{ripple}$$<$$A_{rough}$ is no longer valid.
 Therefore the introduced mechanism is reversible. Furthermore, from theoretical point of view
for an infinitely slow rate of CRM~\cite{hui}, the
relation $\Delta A=\langle W\rangle$ (in common non-equilibrium
simulations) is allowed~\cite{jar}. In this case authors of
Ref.~\cite{1} must report the used infinitely small rate and justify
how it is in practice valid. We show $\Delta A\neq\langle W\rangle$
for the system that was used in Ref.~\cite{1}.

To define the rippled state it was stated in Ref.~\cite{1}:
`\emph{we might state that any primary stress on GS, for
example, in its preparation in the experiments, can construct
ripples that will survive during the experimental measurements}'.
According to this statement, we assume that each GS
(nanoribbons) with boundary stress is in the rippled state
(regardless of the shape of the ripples and the other corresponding
properties).

\begin{figure}
\begin{center}
\includegraphics[width=0.45\linewidth]{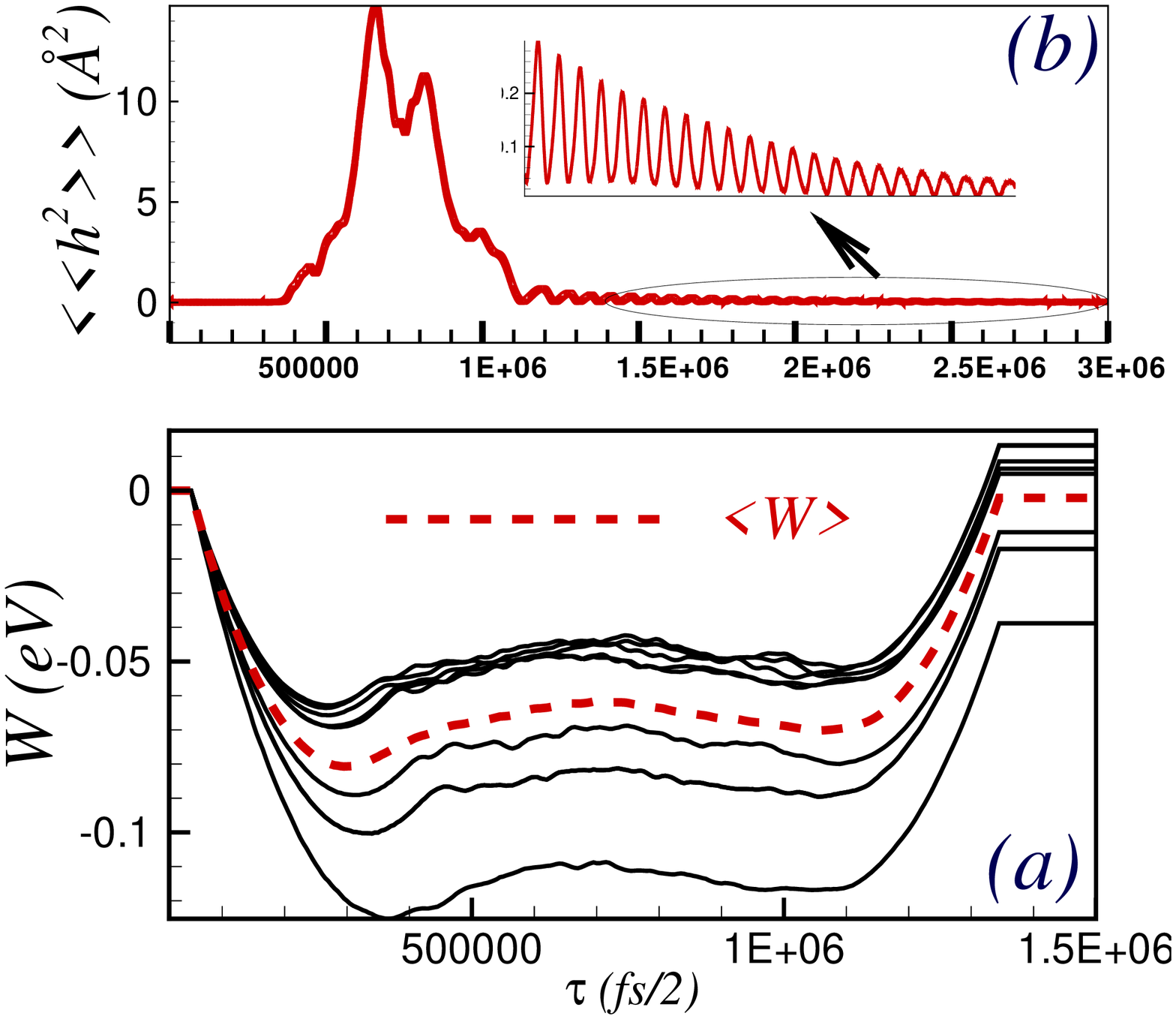}
\includegraphics[width=0.45\linewidth]{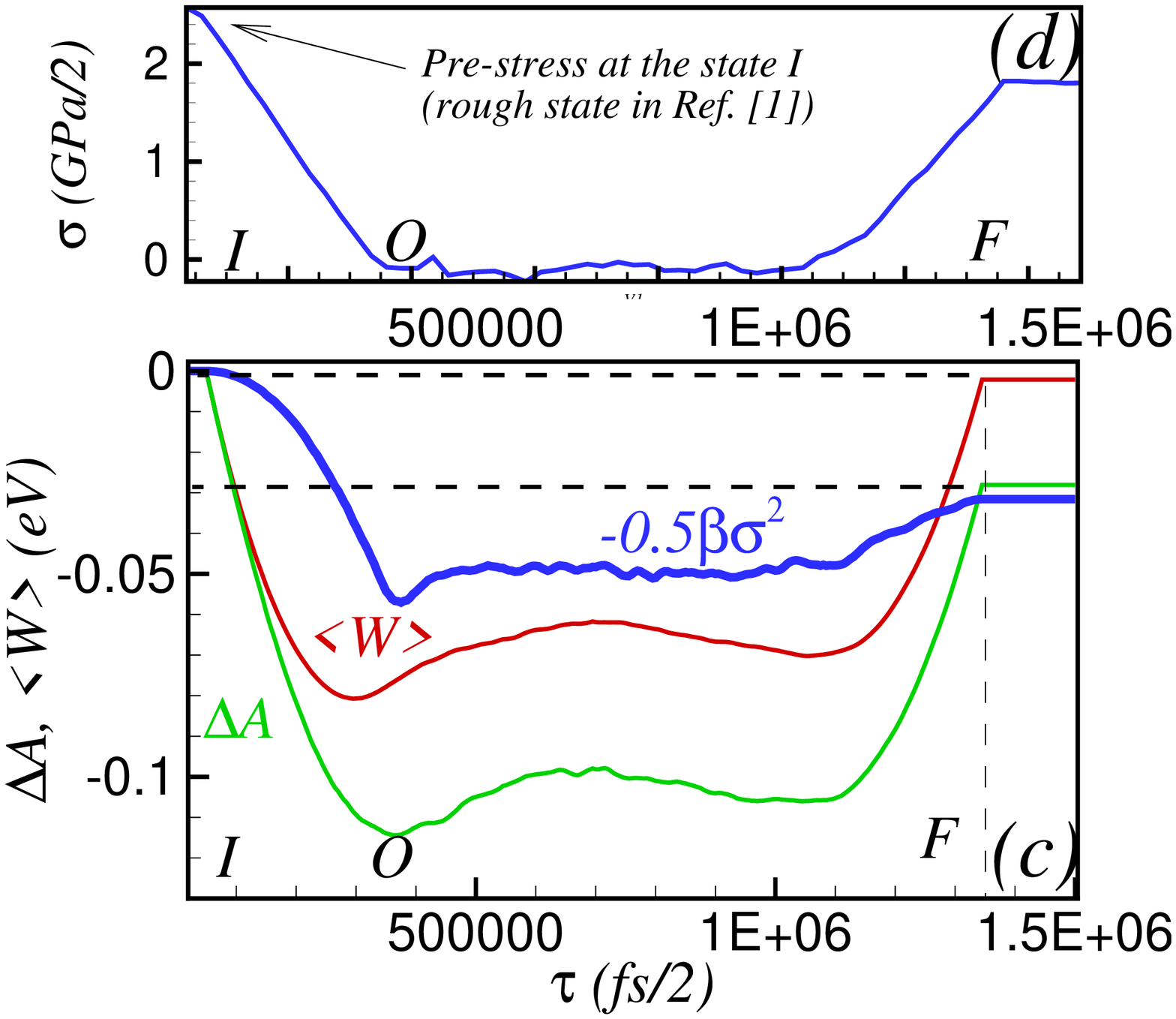}
\caption{(Color online) (a) Variation of the total works for 8
different trajectories (solid curves) and the average work (dashed
curve), (b) $\langle\langle h^2\rangle\rangle$ vs time (inset is zoomed region after
relaxation), (c) free energy change, the average work and dissipated
work and (d) boundary stress
 during CRM.\label{fig1} }
\end{center}
\end{figure}

We simulated the same
sample that was used in Ref.~\cite{1}. A GS with
80$\times$40 atoms by using the same method at T=55~K, the
 strain rate for CRM is 0.0117/ns. The time step is 0.5~fs and
after 25 ps (for thermal equilibrium, i.e. state `\emph{I}'), we compressed the system for 325 ps and then
 relaxed it by moving back to its initial position (state `\emph{F}') for
other 325 ps. Obviously this is done by applying forces on the two rows of atoms at the (longitudinal) ends.
Fig.~\ref{fig1}(a) shows 8 trajectories which are the variation of
the work done on the GS versus time (8 simulations were done
 with different initial conditions)
and the variation of $\langle\langle h^2\rangle\rangle$
(Fig.~\ref{fig1}(b)). The dashed curve is $\langle W\rangle$ which
its variation is related to the dissipated work~\cite{jar}. As we
see~(from Fig.~\ref{fig1}(b))$\langle\langle h^2\rangle\rangle$  (after relaxation) decreases
with time. Therefore the amplitude of the final ripples decreases
after CRM. Fig.~\ref{fig1}(c)~shows the change
in the free energy using Jarzynski equality (i.e. $\Delta
A=-\beta^{-1}\ln \langle{\exp(-\beta W)\rangle}$), $\langle
W\rangle$ and
 $\Delta A-\langle W\rangle(=-0.5\beta \sigma_w^2)$ and
clearly $\Delta A\neq\langle W\rangle$. The difference
$\Delta A-\langle W\rangle$ is just the dissipated work which is
associated with the increase of entropy~\cite{jar,hui}. Practically, the latter difference must be considered even by very small rate of CRM which is in
contrast to the assumption  $\Delta A=\langle W\rangle$ in Ref.~[1].

The size of the system at
state `I' is assumed to be equal to a flat GS (it is due to performing NVT ensemble in Ref. [1]) for a sample size
 with $80 \times 40$ atoms (\emph{`rough state'} in Ref.~\cite{1}). Fig.~\ref{fig1}(d)
 shows the variation of (absolute value of) boundary force per width (i.e. stress) versus CRM time (the averaged value over 8 simulations). As
we see from  the top panel in Fig.~\ref{fig1}(d) at state `\emph{I}'
(`\emph{rough state}' in Ref.~[1]) the system is not free from
stress and it feels pre-stress around  4 GPa which is not
negligible~\cite{lee} (the reason is that the GS does not have
the optimum size~\cite{mity} at `\emph{I}'). Here we emphasize that the behavior
of boundary stress around points `\emph{I}' (and `\emph{F}') is
always similar to that one we presented here.
 After some compression steps the free energy curve shows the first minimum, i.e. state `O'.  At this state
the system is almost free of boundary stress (Fig.~\ref{fig1}(d)).
Moving back to the initial position by
stretching the system, after passing another minimum~\cite{minima}  we obtain the
state `\emph{F}', which is not free from boundary
stress too (`\emph{ripple state}' in Ref.~\cite{1}).
Therefore  both the initial and final state in CRM are rippled states~\cite{mity}.
Moreover, the state `\emph{O}' should be considered as the true `\emph{rough state}' in Ref.~[1], where
there is no boundary stress and the relation $A_{O}-A_{F}<0$
holds~\cite{mity}. The latter inequality implies that the state
`\emph{O}' is much stable than the rippled states `\emph{F}' and `\emph{I}'. In
Ref.~[1] the inequality $A_{F}-A_{I}$ is referred  to as the
difference between rippled (with stress) and the rough state (without stress) i.e. difference between
two dashed horizontal lines in Fig.~\ref{fig1}(c), however  we see
that this is only the difference between two rippled states, i.e.
`\emph{I}' and `\emph{F}'. The final remark is that in Ref.~[1], there is no physical reason given for
 the irreversibility except that $A_{F}-A_{I}<0$, which we have shown this is not valid.

Note that the existence of the boundary stresses at the states `\emph{I}' and `\emph{F}' are
independent of other parameters (e.g. rate for CRM, amount of compression,size of the system and
temperature~\cite{miti2}).

In summary, the true `\emph{rough state}' has lower free energy with
respect to  the rippled state. Using the Jarzynski equality  is
necessary when calculating the free energy change in the
CRM of GS even for very slow
evolution. Both initial and final states in the simulations of
Ref.~[1] are rippled states thus the CRM is reversible.

\small\small{}

\begin{thebibliography}{99}
\bibitem{miti2} We performed another simulation with smaller rate
(0.0064/ns), the smoother curves in Fig.~\ref{fig1} was found, but
the overall behavior is the same.
\end{thebibliography}
\end{document}